\documentclass[qsecnum,amsmath,preprintnumbers,superscriptaddress,nofootinbib,aps,prd,12pt,a4paper]{revtex4-1}
\pdfoutput=1
\usepackage[utf8]{inputenc}
\usepackage{amsmath}
\usepackage{amsfonts}
\usepackage{amssymb}
\usepackage{amsthm}
\usepackage{hyperref}
\usepackage{xcolor}
\usepackage{bm}
\usepackage{graphicx}

 \def\be{\begin{equation}}
\def\ee{\end{equation}}
 \def\ba{\begin{align}}
\def\ea{\end{align}}
\def\bea{\begin{eqnarray}}
\def\eea{\end{eqnarray}}
\def\d{\partial}
\newcommand{\di}{\mathrm d}
\def\a{\alpha}
\def\b{\beta}
\def\g{\gamma}
\def\m{\mu}
\def\n{\nu}
\def\t{\tau}

\def\l{\lambda}

\def\s{\sigma}

\def\D{\Delta}
\def\P{{\cal P}}

\def\bN{\bar{N}}

\newcommand{\bseq}{\begin{subequations}}
\newcommand{\eseq}{\end{subequations}}

\begin{document}

\preprint{CERN-TH-2019-065, INR-TH-2019-007}

\title{{\bf Towards the renormalization group flow of Ho\v rava
    gravity \\ in
$(3+1)$ dimensions}}
\author{Andrei O. Barvinsky}
\address{Theory Department, Lebedev Physics Institute \\ Leninsky Prospect 53, Moscow 117924, Russia}
\author{Mario Herrero-Valea}
\address{Institute of Physics, Laboratory of Particle Physics and Cosmology, \'Ecole Polytechnique F\'ed\'erale de Lausanne\\ CH-1015 Lausanne, Switzerland}
\author{Sergey M. Sibiryakov}
\address{Institute of Physics, Laboratory of Particle Physics and Cosmology, \'Ecole Polytechnique F\'ed\'erale de Lausanne\\ CH-1015 Lausanne, Switzerland}
\address{Theory Department, CERN \\
1 Esplanade des Particules, CH-1211 Gen\`eve 23, Switzerland}
\address{Institute for Nuclear Research of the Russian Academy of Sciences
60th October Anniversary Prospect, 7a, 117312 Moscow, Russia}

\begin{abstract}
We compute the renormalization group running of the Newton constant
and the parameter $\lambda$ in $(3+1)$-dimensional projectable 
Ho\v rava gravity. We use the background field method expanding around
configurations with flat spatial metric, but non-vanishing shift. This
allows us to reduce the number of interaction vertices and thereby
drastically simplify the calculations. The gauge invariant
$\b$-function of $\l$ has two families of zeros, attractive in the
infrared and ultraviolet respectively. They are candidates for the
fixed points of the full renormalization group flow of the theory, once the
$\b$-functions for the rest of the couplings are added.  
\end{abstract}
\maketitle
\newpage
\section{Introduction}

Since the formulation of general relativity (GR) a century ago and of
quantum mechanics a few years later the quest for a theory unifying
the two --- quantum gravity (QG) --- has been one of the biggest
endeavors in theoretical physics. While other known fundamental
interactions are successfully described within the formalism of local
perturbative quantum field theory (QFT), most approaches to QG
involve departures from this framework. Despite a number of
impressive achievement of these research programs, the quest for QG remains
largely unaccomplished. Thus, it is reasonable to wonder if there is
still a way to formulate QG within the realm of QFT. 
 
However, a straightforward approach to this goal fails due to the
dimensionful character of Newton's constant $G$ in dimensions greater
than two, in particular in $(3+1)$ dimensions. This renders the theory
non-renormalizable, with the number of divergences increasing at every
loop order and requiring an infinite set of counterterms suppressed
by higher powers of the Planck mass $M_P\sim \sqrt{G^{-1}}\sim 10^{19}
{\rm GeV}$. Thus, perturbative GR can be
thought at most as an effective low-energy theory of gravitational
interactions, see e.g. \cite{Donoghue:1994dn,Burgess:2003jk,Donoghue:2017pgk}, 
unable to describe
physics above the ultraviolet (UV) cutoff given by $M_P$. 

A simple way out of this problem in four space-time dimensions is to
extend the gravitational Lagrangian with terms quadratic in curvature
\cite{Stelle:1976gc,Fradkin:1981hx,Avramidi:1985ki} so
that UV divergences are softened by higher powers of four-momenta in
the propagators. Although this theory is thereby renormalizable and
even asymptotically free for some regions of the parameter space,
unitarity is jeopardized by the presence of four time derivatives in
the action, which gives rise to a propagating Ostrogradsky ghost
with negative norm in the Hilbert space~\cite{Stelle:1977ry,Woodard:2015zca}.  

It was pointed out by P. Ho\v rava \cite{Horava:2009uw}
that unitarity can be preserved in this kind of models if we sacrifice
Lorentz invariance (LI) at very high energies. This allows one to
introduce into the Lagrangian extra higher spatial derivatives which
will soften UV divergences while at the same time keeping only two
time derivatives, so that the ghosts are avoided. For
this proposal to work, one must supplement the spacetime manifold with
a preferred foliation into spatial slices and a privileged time
direction. 
Then it is possible to formulate a theory containing only marginal
operators and relevant deformations with respect to an anisotropic
(Lifshitz) scaling in $(d+1)$ dimensions\footnote{Throughout the text we
  use 
Latin indices to denote the spatial
  directions, $i=1,\dots,d$.}, 
\begin{align}\label{eq:scaling}
t\rightarrow b^{-d}t,\qquad x^i\rightarrow b^{-1}x^i\;,
\end{align}
instead of the usual isotropic scaling. As a consequence, a theory of
gravity formulated under this setting cannot be LI nor completely
diffeomorphism invariant. Instead, it will enjoy a restricted symmetry
in terms of foliation preserving diffeomorphisms (FDiff), 
\begin{align}
t\rightarrow t'(t),\qquad x^{i}\rightarrow x'^i(t,x)\;,
\end{align}
where $t'(t)$ must be a monotonic function. That is, we are left with
invariance under time reparameterizations and spatial time-dependent
diffeomorphisms. 

Ho\v rava's proposal gave rise to numerous follow-up publications
studying different aspects of Ho\v rava gravity (HG) (see
\cite{Mukohyama:2010xz,Sotiriou:2010wn,
  Blas:2018flu,Wang:2017brl} for reviews). These include an exhaustive
study of the low energy regime of the theory, which led to the
identification of a version of HG -- the healthy non-projectable model
\cite{Blas:2009qj} -- which can reproduce the known phenomenology of
GR at the scales in which the latter has been tested
\cite{Blas:2014aca}. While its parameter space has been strongly
constrained by the observation of a gravitational wave signal from a
binary neutron star merger \cite{Monitor:2017mdv}, it still remains
phenomenologically viable \cite{Gumrukcuoglu:2017ijh}. In this theory, a
certain amount of LI violation persists at all scales
\cite{Blas:2010hb}. This can lead to interesting cosmological
applications 
\cite{Blas:2011en,Audren:2013dwa}. However, it also demands the
presence of a mechanism\footnote{For suggestions of such mechanism see
\cite{Chadha:1982qq,Anber:2011xf,Sundrum:2011ic,
Kiritsis:2012ta,Bednik:2013nxa,Kharuk:2015wga,
GrootNibbelink:2004za,Bolokhov:2005cj}.} 
suppressing the percolation of LI violation
to the visible matter sector, where LI has been tested to a very high
precision~\cite{Liberati:2013xla}. 

Regarding renormalizability, the situation is less clear. Although the
Lagrangian of the non-projectable model is power-counting
renormalizable, the lapse element of the metric induces an
instantaneous interaction \cite{Blas:2010hb,Blas:2011ni} 
which can potentially lead to
non-local loop divergences \cite{Barvinsky:2015kil}. 
Whether this indeed happens or not
is an open question.

There has been a significant progress in understanding the quantum dynamics
of a reduced version of HG known as the projectable model, where the
lapse $N$ is restricted to be only time-dependent. Then it can be
algebraically fixed to take a given value, say $N=1$. In this way one 
gets rid of time
reparameterization invariance as a symmetry of the theory, being left
only with spatial diffeomorphisms. 
This version of the theory does not have a stable homogeneous vacuum
with flat Minkowski metric and thus is unlikely to reproduce realistic
phenomenology, at least in the controllable regime of weak coupling
\cite{Blas:2009yd,Koyama:2009hc,Blas:2010hb} (see, however,
\cite{Mukohyama:2010xz,Izumi:2011eh,Gumrukcuoglu:2011ef}). 
Nevertheless, it captures several salient
features of QG, such as propagating spin-2 gravitons (in spacetime
dimensions greater than 3) and a large gauge group of coordinate
transformations precluding existence of local gauge-invariant
observables. 
The
projectable model has been proven to be renormalizable in the strict
sense, 
i.e.
all divergences are proportional to gauge invariant operators already
present in the bare action, in any space-time 
dimension~\cite{Barvinsky:2015kil,Barvinsky:2017zlx}. 

Furthermore, the computation of the full renormalization group (RG) 
flow of the theory in $(2+1)$ spacetime dimensions reveals the
presence of an asymptotically free UV fixed
point~\cite{Barvinsky:2017kob}. Unlike $(2+1)$-dimensional GR, this
theory possesses a propagating scalar degree of freedom and thus
provides an example of a perturbatively UV complete
gravitational field theory with non-trivial local dynamics.  
Other results along these lines include the study of a conformal
reduction of the $(2+1)$-dimensional theory \cite{Benedetti:2013pya}, 
the computation of the anomalous dimension of the cosmological
constant \cite{Griffin:2017wvh}, also in $(2+1)$ dimensions, 
and the contribution of scalar matter 
into the renormalization of the gravitational
couplings~\cite{DOdorico:2014tyh}. 

In this paper we make a step towards uncovering the RG flow of HG in
$(3+1)$ dimensions. This case differs from the $(2+1)$-dimensional
model in two respects. First, the physical spectrum of the theory now
includes a spin-2 mode --- graviton. Second, on the technical side,
the structure of the propagators and vertices is significantly more
complicated than in $(2+1)$-dimensional HG, which makes a direct
application of the approach used in \cite{Barvinsky:2017kob}
infeasible. We find a way to modify this approach that allows us to
compute the 1-loop $\b$-functions of the two couplings entering the
kinetic part of the action. Renormalization of the 
potential part requires different techniques and will be reported
elsewhere. 

The paper is organized as follows. In Sec.~\ref{sec:2} we review the
structure of projectable HG and its relevant properties. In
Sec.~\ref{sec:3} we describe our calculation. In
Sec.~\ref{sec:results} we report the results for the $\b$-functions
and discuss their implications. Section~\ref{sec:conclusions} is
devoted to conclusions. Appendices provide additional details.

\section{Projectable Ho\v rava Gravity}
\label{sec:2}
Ho\v rava gravity \cite{Horava:2009uw} is formulated by considering
the spacetime metric together with a preferred foliation along the
time direction. This induces an Arnowitt--Deser--Misner (ADM)
decomposition of the metric, 
\begin{align}
ds^2 = N^2 \di t^2 -\gamma_{ij}(\di x^i + N^i \di t)(\di x^j+N^j \di t)\;.
\end{align}
The lapse $N$, the shift $N^i$ and the spatial metric $\g_{ij}$ on the
foliation slices transform in the standard way under FDiff
transformations, 
\begin{align}
\label{FDiff}
N\mapsto N\frac{\di t}{\di t'},\qquad  
N^i\mapsto \left(N^j \frac{\partial x'^i}{\partial x^j} -
  \frac{\partial x'^i}{\partial t}\right)\frac{\di t}{\di t'},\qquad 
\gamma_{ij}\mapsto \gamma_{kl}\frac{\partial x^k}{\partial
  x'^i}\frac{\partial x^l}{\partial x'^j}\;. 
\end{align}
Their dimensions under the anisotropic scaling \eqref{eq:scaling}
are,\footnote{We say that a field $\Phi$ has dimension $[\Phi]$ if it
  transforms under \eqref{eq:scaling} as
$\Phi\mapsto b^{[\Phi]}\Phi$.
Throughout the paper we will understand the notion of dimension in
this sense.
}
\begin{align}
[N]=[\gamma_{ij}]=0,\qquad [N^i]=d-1\;.
\end{align}
The FDiff symmetry, together with time-reversal invariance, parity and
the requirement of power-counting renormalizability fix the action of HG to be 
\begin{align}
\label{HGact}
S=\frac{1}{2G}\int \di t \di^d x \sqrt{\gamma}\, N
\left(K_{ij}K^{ij}-\lambda K^2 -{\cal V}\right)\;. 
\end{align}
Here $G$ and $\lambda$ are dimensionless coupling constants and the extrinsic curvature of the foliation is given by
\begin{align}
K_{ij}=\frac{1}{2N}\left(\dot{\gamma}_{ij}-\nabla_i N_j -\nabla_j N_i\right)\;,
\end{align}
where dot denotes time derivative and $\nabla_i$ is the covariant
derivative associated to the spatial metric; $K\equiv
K_{ij}\gamma^{ij}$ 
denotes the trace of $K_{ij}$.
FDiff transformations (\ref{FDiff}) are compatible with an assumption
of the lapse function being constant on the foliation slices,
$N=N(t)$. This specifies the projectable version of the model
\cite{Horava:2009uw}, to which we restrict in what follows. 
In this case, we can set $N=1$ as an algebraic gauge fixing for time
reparameterization invariance, eliminating it from the set of dynamical
fields and leaving us only with spatial diffeomorphisms as a gauge
symmetry. 

The potential ${\cal V}$ contains all possible marginal and relevant
operators\footnote{I.e. operators with dimension less or equal than the
  absolute value of the dimension of the spacetime integration measure
$\di t\di^d x$, which is equal to $2d$.} 
invariant under FDiff and without time derivatives. In
$d=3$ and after using Bianchi identities and integration by parts,
these reduce to \cite{Sotiriou:2009gy},
\begin{align}
\nonumber {\cal V}&=2\Lambda-\eta R +\mu_1 R^2 +\mu_2 R_{ij}R^{ij}\\
&+\nu_1 R^3 +\nu_2 R R_{ij}R^{ij}+\nu_3 R^i_j R^j_k R^k_i+\nu_4
\nabla_i R \nabla^i R +\nu_5 \nabla_i R_{jk} \nabla^i R^{jk}\;, 
\label{HGpot}
\end{align}
where $R_{ij}$ is the Ricci tensor corresponding to the spatial metric
and we have used the fact that in three dimensions the Riemann tensor
is expressed in terms of $R_{ij}$.

The spectrum of perturbations propagated by this action contains a
transverse-traceless (tt) graviton and an additional scalar mode
\cite{Blas:2009yd,Barvinsky:2015kil}. 
Both modes have positive-definite kinetic terms as long as $G>0$ and
$\lambda \in \left(-\infty,\frac{1}{3}\right)\cup (1,\infty)$ and the
theory admits unitary quantization.
Their dispersion relations around a flat background\footnote{By this
  we mean the
  configuration $N^i=0$, $\g_{ij}=\delta_{ij}$. It is a solution of
  equations following from (\ref{HGact}), (\ref{HGpot}) provided the
  cosmological constant $\Lambda$ is set to zero.} are
\bseq
\label{eq:dispersion}
\begin{align}
&\omega^2_{tt}=\eta k^2+\m_2k^4+\n_5k^6\;,\\
&\omega^2_s=\frac{1-\l}{1-3\l}\big(-\eta
k^2+(8\m_1+3\m_2)k^4\big)+\nu_s k^6\;,
\label{dispscalar}
\end{align}
\eseq
where $k$ is the spatial momentum and we have defined
\begin{align}
\label{nus}
\nu_s=\frac{(1-\l)(8\n_4+3\n_5)}{1-3\l}\;.
\end{align}
A problem arises in the low-energy limit, where the dispersions
relations are dominated by the terms proportional to $k^2$. Due to the
negative sign in front of this term, the scalar mode 
behaves as a tachyon at low energies, implying that flat space is not a
stable vacuum of the theory. Attempts to suppress the instability by
choosing $\l$ close to 1 lead to the loss of perturbative control
\cite{Blas:2009yd,Koyama:2009hc,Blas:2010hb} (see, however, 
\cite{Mukohyama:2010xz,Izumi:2011eh,Gumrukcuoglu:2011ef} for
suggestions to restore the control by rearranging the perturbation
theory at the classical level). Alternatively, the instability can be
eliminated by tuning $\eta=0$ or by expanding around a curved
vacuum. In both cases the theory does not appear to reproduce GR in
the low-energy limit, as there is no regime where the dispersion
relation of the tt-graviton would have the relativistic form
$\omega^2_{tt}\propto k^2$. On the other hand, both dispersion
relations (\ref{eq:dispersion}) are perfectly regular at high momenta
for $\nu_5,\nu_s>0$. This is the relevant region for the
structure of UV divergences which are the focus of the present paper. 

For the rest of this work we restrict to the sector of marginal
operators that dominate the UV behavior of the theory by 
setting $\eta=\mu_1=\mu_2=0$. We also perform a Wick rotation to
``Euclidean'' time $\tau$,
which amounts to flipping the sign of the potential term. Thus we
consider the action,
\begin{align}\label{eq:action}
\nonumber S=\frac{1}{2G}\int \di \t \di^3x
\sqrt{\gamma}\,&\left(K_{ij}K^{ij}-\lambda K^2+\nu_1 R^3 +\nu_2 R
  R_{ij}R^{ij}\right.\\ 
&\left.+\nu_3 R^i_j R^j_k R^k_i +\nu_4 \nabla_i R \nabla^i R +\nu_5
  \nabla_i R_{jk} \nabla^i R^{jk}\right) \;.
\end{align}
We are interested in the 1-loop correction to this action and the
corresponding RG flow of the couplings $\{G,\l,\nu_1,\ldots,\nu_5\}$.

\section{The background-field approach}
\label{sec:3}

In our calculation we use the background field method
\cite{Abbott:1981ke}. 
We expand the fields about arbitrary background values,
\begin{align}
\label{backgr1}
\gamma_{ij}=\bar{\gamma}_{ij}+h_{ij},\qquad N_i=\bar{N}_i+n_i\;,
\end{align}
and expand the action (\ref{eq:action}) up to second order in the
fluctuations $h_{ij},n_i$. One-loop divergences will be captured as
contributions to operators carrying background fields after
integrating out the fluctuations. 
This procedure should work in general and, similarly to the
case in $(2+1)$  dimensions, we should be able to obtain the
contributions to the whole set of coupling constants
$\{G,\lambda,\nu_a\}$ by an appropriate choice of background.
However, the complexity of computations quickly grows with the number
of space-time dimensions. This stems mainly from the increased number
of terms in the potential part of the action and their more
complicated structure. Indeed,
in $(2+1)$ dimensions there is a single term $R^2$ which generates
about twenty different contributions when expanded about an arbitrary
background (\ref{backgr1}). 
In $(3+1)$ dimensions we have five operators shown in
\eqref{eq:action} 
which generate hundreds of structures. This makes a
complete calculation of the RG flow in any dimension higher than
$(2+1)$ highly challenging. In particular, it does not appear feasible
to apply the diagrammatic approach of \cite{Barvinsky:2017kob} to
renormalization of the operators in the potential. Rather, one will
have to use the covariant techniques developed in
\cite{Barvinsky:1985an,DOdorico:2015pil,Barvinsky:2017mal} 
or other
advanced tools. 

Fortunately, the situation is different for the operators in the
kinetic part of the action, i.e. the two operators involving the
extrinsic curvature. We observe that the latter contains the shift
vector $N^i$ in a specific combination. Then, exploiting the known
gauge invariance of divergences
\cite{Barvinsky:2015kil,Barvinsky:2017zlx}, we can extract the
renormalization of $G$ and $\l$ not only from the terms with time
derivatives of the spatial metric as in \cite{Barvinsky:2017kob}, 
but also from terms with spatial gradients of $N^i$.
Since the shift only enters into the kinetic term, all interaction
vertices contributing to the renormalization of these operators will
also come exclusively from the expansion of the kinetic term.
The potential, as we will see shortly, only contributes into the
propagators.  
Furthermore, we can restrict to backgrounds with flat spatial metric
$\bar\g_{ij}=\delta_{ij}$, as long as we allow the background shift
$\bar N^i$ to take arbitrary space-dependent values.
All this together means that we can derive the renormalization of $G$
and $\lambda$ avoiding the complexity mentioned above and allows us
to carry out the computation in a reasonably simple way.

According to the above discussion, we consider a family of backgrounds
characterized by flat spatial metric and an arbitrary space-dependent
shift. Thus, we specify (\ref{backgr1}) to 
\begin{align}
\label{flatbg}
\gamma_{ij}=\delta_{ij}+h_{ij}(\t,x),\qquad N_i=\bar{N}_i(x)+n_i(\t,x)\;.
\end{align}
The Lagrangian is then expanded to quadratic order in $h_{ij}$ and
$n_i$. Note that we define the perturbation of the shift with lower
indices: this choice turns out to be more convenient as it reduces the
number of terms in the perturbed Lagrangian. The perturbation of the
component with the upper indices is then given by 
\be
\label{deltaNup}
\delta N^i=n^i-h^i_j\bar N^j\;,
\ee
where on the r.h.s. the indices are raised using the flat background
metric $\delta^{ij}$. From now on we will not distinguish upper and lower
indices in the expressions expanded around the background
(\ref{flatbg}); all repeated spatial indices will be simply
summed over.  

We are interested in those operators in the 
effective action for the background fields that contain the background
shift. 
With our choice of flat background spatial metric, these are
\begin{align}\label{eq:lagrangian_bg}
S_{\bar{N}}=\frac{1}{2G}\int \di \t \di^3x \, \left(\frac{1}{2}\partial_i
  \bar{N}_j \partial_i
  \bar{N}_j+\left(\frac{1}{2}-\lambda\right)\partial_i
  \bar{N}_i\partial_j \bar{N}_j \right) \;.
\end{align}
By computing the one-loop contribution to these operators we will be
able to capture the renormalization of $G$ and $\lambda$. 

\subsection{Covariant gauge fixing}
\label{sec:gaugefix}

The action \eqref{eq:action} is invariant under the residual part of
FDiff that preserves the condition $N=1$, namely time-dependent
spatial diffeomorphism of the form 
\begin{align}
x^{i}\rightarrow x'^i(\t,x)\;.
\end{align}
In order to proceed with the quantization of the theory we must thus
append the bare action with a gauge fixing term for this symmetry.
Within the background-field approach, the latter is chosen to be
invariant under the {\it background gauge transformations} that in our
case read,
\be
\label{bgFDiff}
\bar N^i\mapsto \bar N^j\frac{\d x'^i}{\d x^j}-\frac{\d x'^i}{\d \t},\qquad
\bar\g_{ij}\mapsto\bar \gamma_{kl}\frac{\partial x^k}{\partial
  x'^i}\frac{\partial x^l}{\partial x'^j}\;.
\ee
This ensures that the one-loop background effective action 
consists of gauge invariant operators and allows one to extract the
renormalized couplings as the coefficients in front of such operators
\cite{Abbott:1981ke}. By comparing (\ref{bgFDiff}) to (\ref{FDiff}) it
is straightforward to see that the shift perturbation with upper index
$\delta N^i\equiv N^i-\bar N^i$ and the metric perturbation $h_{ij}$ transform
as a contravariant vector and a covariant tensor
respectively\footnote{Instead, the perturbation of the shift with the
  lower index $n_i$, which we take as our primary variable, transforms
inhomogeneously.}. They serve as the building blocks to construct the
gauge fixing term.  

We
choose the gauge fixing introduced in \cite{Barvinsky:2015kil} which
ensures that 
the propagators of all fluctuations decay uniformly in frequency
and momentum in a way compatible with the anisotropic scaling.
It reads, 
\begin{align}
S_{\rm gf}=\frac{\sigma}{2G}\int \di \t \di^3x\, F^i
{\cal O}_{ij} F^j\;, 
\end{align}
where
\bseq
\begin{align}
&F^i=D_\t (\delta N^i)+\varsigma \bar{K}^{i}_{j}\,\delta N^j
+\frac{1}{2\sigma}\left({\cal O}^{-1}\right)^{ij}\left(\d_k h^k_j 
-\lambda \d_j h\right)\;,\\
&\left({\cal O}^{-1}\right)^{ij}=\delta^{ij}\Delta^2 
+ \xi \d^i \Delta\d^j\;.
\end{align}
\eseq
Here $h\equiv h_i^i$ is the trace of the metric fluctuations, 
$\D\equiv \d_i\d^i$ is the flat-space Laplacian, $\sigma$ and $\xi$
are free parameters and 
\be
\label{DtdN}
D_\t (\delta N^i)=\delta\dot{N}^i -\bar{N}^k\, \d_k \delta N^i
+\d_k \bar{N}^i\, \delta N^k\;
\ee
is the covariant time-derivative of $\delta N^i$.
Note that we have added to the gauge-fixing function a contribution
proportional to the background extrinsic curvature $\bar K^i_j$ 
with an arbitrary
coefficient $\varsigma$. This additional 
freedom, which was not present in the original gauge fixing of
\cite{Barvinsky:2015kil}, can be used to simplify the Lagrangian for
perturbations and as an extra check of gauge invariance of our
results. 

The presence of the non-local operator ${\cal O}_{ij}$ leads to appearance
of non-local terms in the gauge-fixing action that are quadratic in
$\delta N^i$. Following \cite{Barvinsky:2015kil,Barvinsky:2017kob} we
eliminate this non-locality by integrating {\it in} an extra field
$\pi_i$, so that we rewrite,
\be
\begin{split}
&\frac{\sigma}{2G}\big(D_\t(\delta N^i)+\varsigma \bar{K}^i_j\,\delta
N^j\big)
\,{\cal O}_{ik}\,
\big(D_\t(\delta N^k)+\varsigma \bar{K}^{k}_{l}\delta N^l\big)\\
&\mapsto
\frac{1}{2G}\left[ \frac{1}{\sigma}\pi_i \left({\cal
      O}^{-1}\right)^{ij}\pi_j-2i \pi_i \big(D_\t(\delta N^i)+\varsigma
  \bar{K}^{i}_{j}\delta N^j\big)\right]\;.
\end{split}
\ee
Then the whole gauge-fixing action takes the local form, 
\be
\begin{split}
S_{\rm gf}=&\frac{1}{2G}\int\di\t\di^3x\,\left[ \frac{1}{\sigma}\pi_i
  \left({\cal O}^{-1}\right)^{ij}\pi_j-2i \pi_i \big(D_\t(\delta
  N^i)+\varsigma 
\bar{K}^i_j\,\delta N^j\big)\right.\\
&\left.+\big(D_\t(\delta N^i)+\varsigma 
\bar{K}^i_j\,\delta N^j\big)\big(\d_k h^k_i-\lambda \d_i h\big)
+\frac{1}{4\sigma}\big(\d_k h^k_i-\lambda \d_i h\big)
\left({\cal O}^{-1}\right)^{ij}\big(\d_l h^l_j-\lambda \d_j
h\big)\right].
\label{Sgf}
\end{split}
\ee

Finally, we have to add the anticommuting Faddeev-Popov ghost $c^i$ and
antighost $\bar c_i$. The action for them is constructed 
out of the gauge fixing function as
\begin{align}
\label{Sgh}
S_{\rm gh}=-\frac{1}{G}\int \di \t \di^3 x \,\bar{c}_i \mathbf{s} F^i\;,
\end{align}
where the BRST operator $\mathbf{s}$ can be understood as implementing
a gauge transformation with the gauge parameter replaced by the ghost,
\bseq
\begin{align}
&\mathbf{s}h_{ij}=\d_i c_j +\d_j c_i +\d_i c^k h_{jk}+\d_j c^k h_{ik}
+c^k\d_k h_{ij}\;,\\
&\mathbf{s}(\delta N^i)=D_\t c^i -\delta N^j \d_j c^i +c^j \d_j \delta N^i\;.
\end{align}
\eseq
To sum up, the total action that we are going to work with reads,
\be
\label{Stot}
S_{\rm tot}=S^{(2)}+S_{\rm gf}+S_{\rm gh}\;,
\ee
where $S^{(2)}$ is the result of expanding (\ref{eq:action}) to
quadratic order in $h_{ij},n_i$ and $S_{\rm gf}$, $S_{\rm gh}$ are
given by (\ref{Sgf}), (\ref{Sgh}) respectively.

\subsection{Propagators}
To set up the computation along the lines of \cite{Barvinsky:2017kob}, 
we need to derive the propagators for the 
dynamical field $\{h_{ij}, n_i, \pi_j, c_i,\bar c_j\}$. They are 
obtained by setting the background shift $\bar N_i$ to zero in the
total quadratic action for the fluctuations (\ref{Stot}). The
resulting Lagrangian splits into several pieces, 
\bseq
\label{L2}
\begin{align}
{\cal L}_{h}=\frac{1}{2G}\bigg[&\frac{1}{4}\dot h_{ij}^2-\frac{\l}{4}\dot
h^2
-\frac{\nu_5}{4}h_{ij}\D^3 h_{ij}
+\bigg(-\frac{\nu_5}{2}+\frac{1}{4\s}\bigg)\d_k h_{ki}\,\D^2\d_l h_{li}\notag\\
&+\bigg(-\nu_4-\frac{\nu_5}{2}-\frac{\xi}{4\s}\bigg)
\d_i\d_k h_{ik}\,\D\d_j\d_l h_{jl}
+\bigg(2\nu_4+\frac{\nu_5}{2}+\frac{\l(1+\xi)}{2\s}\bigg)
\D^2h\,\d_i\d_kh_{ik}\notag\\
&+\bigg(-\nu_4-\frac{\nu_5}{4}-\frac{\l^2(1+\xi)}{4\s}\bigg)
h\D^3 h\bigg],
\label{L2hh}\\
\label{L2nn}
{\cal L}_{n}=\frac{1}{2G}\bigg[&\frac{1}{\s}\pi_i\D^2\pi_i
-\frac{\xi}{\s}\d_i\pi_i\,\D\d_j\pi_j-2i\pi_i\dot n_i
-\frac{1}{2}n_i\D n_i+\bigg(\frac{1}{2}-\l\bigg)(\d_in_i)^2\bigg],\\
\label{L2cc}
{\cal L}_{\bar c}=-\frac{1}{G}\bigg[&\bar c_i\ddot c_i
+\frac{1}{2\s}\bar c_i\D^3 c_i-\frac{1+2\xi-2\l(1+\xi)}{2\s}
\d_i\bar c_i\,\D^2\d_k c_k\bigg].
\end{align}
\eseq
In deriving these expressions we used integration by parts.
Note that our choice of the gauge fixing ensures cancellation of the
terms mixing $h_{ij}$ and $n_i$ in the expansion of the bare
Lagrangian \cite{Barvinsky:2015kil}.
Inversion of the differential operators appearing in (\ref{L2}) yields
the propagators, 
\bseq
\label{props}
\begin{align}
&\langle h_{ij}h_{kl}\rangle=G\bigg[
2(\delta_{ik}\delta_{jl}+\delta_{il}\delta_{jk})\P_{tt}
-2\delta_{ij}\delta_{kl}\bigg(\P_{tt}-\frac{1-\l}{1-3\l}\P_s\bigg)\notag\\
&\qquad\qquad-2(\delta_{ik}\hat k_j\hat k_l+\delta_{il}\hat k_j\hat k_k
+\delta_{jk}\hat k_i\hat k_l+\delta_{jl}\hat k_i\hat k_k)(\P_{tt}-\P_1)
+2(\delta_{ij}\hat k_k\hat k_l+\hat k_i\hat
k_j\delta_{kl})(\P_{tt}-\P_s)\notag\\
&\qquad\qquad+2\hat k_i\hat k_j\hat k_k\hat k_l
\bigg(\P_{tt}+\frac{1-3\l}{1-\l}\P_s-4\P_1+\frac{2}{1-\l}\P_2\bigg)\bigg]
\;,\label{prophh}
\\
\label{propnn}
&\langle
n_in_j\rangle=G\bigg[\frac{k^2}{\s}(k^2\delta_{ij}-k_ik_j){\cal P}_1
+\frac{1+\xi}{\s}k^2k_ik_j{\cal P}_2\bigg]\;,\\
\label{proppin}
&\langle\pi_i n_j\rangle=G\big[\omega\delta_{ij}{\cal P}_1
+\omega\hat k_i\hat k_j({\cal P}_2-{\cal P}_1)\big]\;,\\
\label{proppipi}
&\langle\pi_i\pi_j\rangle=G\bigg[
\frac{1}{2}(k^2\delta_{ij}-k_ik_j){\cal P}_1+(1-\l)k_ik_j{\cal P}_2\bigg]\;,\\
\label{propcc}
&\langle \bar c_i c_j\rangle=G\big[\delta_{ij}{\cal P}_1
+\hat k_i\hat k_j({\cal P}_2-{\cal P}_1)\big]\;,
\end{align}
\eseq
where $\hat k_i\equiv k_i/k$ is 
the unit vector along the momentum $k_i$, and the
pole structures are given by,
\bseq
\label{poles}
\begin{align}
\label{Pt}
&\P_{tt}=\big[\omega^2+\nu_5 k^6\big]^{-1}\;,\\
\label{Ps}
&\P_s=\bigg[\omega^2+\nu_s k^6\bigg]^{-1}\;,\\
\label{P1}
&{\cal P}_1=\bigg[\omega^2+\frac{k^6}{2\s}\bigg]^{-1}\;,\\
\label{P2}
&{\cal P}_2=\bigg[\omega^2+\frac{(1-\l)(1+\xi)}{\s}k^6\bigg]^{-1}\;.
\end{align}
\eseq
These expressions coincide with those derived in
\cite{Barvinsky:2015kil}. Note that the first two poles correspond to
the dispersion relation of the transverse-traceless and scalar gravitons
of the physical spectrum of the theory, while the rest are gauge
dependent and their positions can be tuned by changing the gauge
parameters 
$\sigma$ and $\xi$.

\subsection{Vertices}

\begin{figure}
\includegraphics[scale=1]{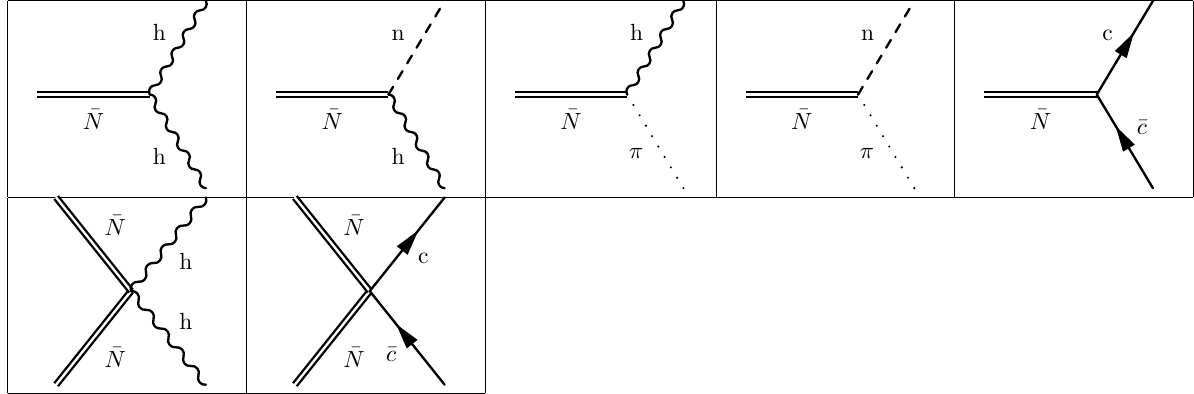} 
\caption{The vertices contributing to the renormalization of
  shift-dependent operators at one-loop.}\label{fig:vertices} 
\end{figure}

The second ingredient needed for the computation are vertices
describing the interaction of external $\bar N^i$-legs with the
dynamical fields. The relevant vertices are depicted in 
Fig.~\ref{fig:vertices} and are obtained by
expanding the Lagrangian up to the given order in the number of
fields. The corresponding formulas are rather lengthy and are
relegated to the Appendix~\ref{app:Lint}.

The most important advantage of focusing on the renormalization of
shift-dependent operators is that all vertices come exclusively
from the kinetic terms and are completely independent of the structure
of the potential, as well as of the gauge parameters $\sigma$ and
$\xi$. Interestingly, this implies that the vertices are universal for
HG in any spacetime dimensions. Another implication is that they are
independent of the couplings $\nu_1,\nu_2,\nu_3$. As the propagators
(\ref{props}) also do not contain these couplings, we conclude that the
one-loop renormalization of $G$ and $\l$ is insensitive to them
altogether.

\subsection{One-loop diagrams}
\begin{figure}
\includegraphics[scale=1]{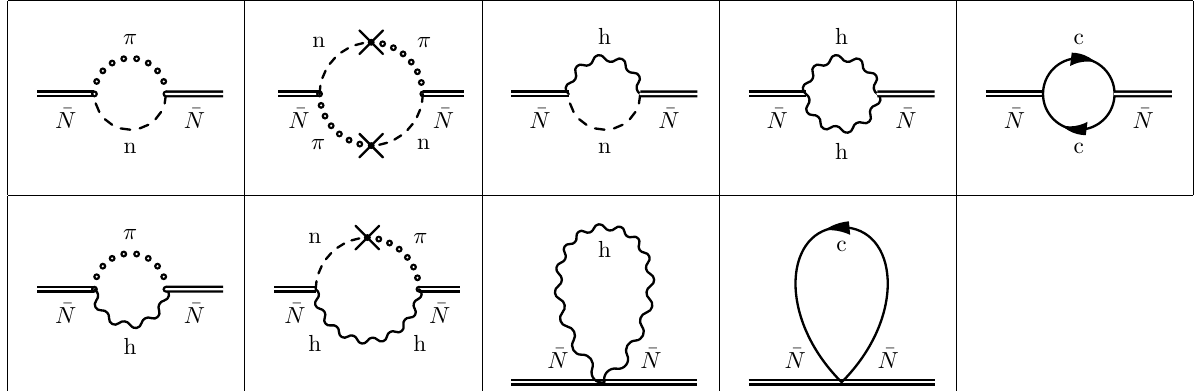} 
\caption{Logarithmically divergent Feynman diagrams contributing to the renormalization of $G$ and $\lambda$ at one-loop level.}\label{fig:diagrams}
\end{figure}

We are interested in capturing the renormalization of the operators 
$\partial_i\bN_j \partial_i \bN_j$ and 
$(\partial_i \bN_i)^2$, which in momentum space read 
$p_ip_i\bN_j\bN_j$, $p_ip_j\bN_i\bN_j$, where $p_i$ denotes the
momentum flowing through the $\bN_i$-legs.
The diagrams contributing to these operators 
are shown in Fig.~\ref{fig:diagrams}. The prototypical form of the
integrals appearing in the fish diagrams (upper row and first two
diagrams in the lower row in the figure) is 
\begin{align}
\label{fishInt}
\int \frac{\di\omega \di^3k}{(2\pi)^4}\,
\Pi_1(\omega,{\bf k})\, \Pi_2(-\omega,-{\bf k}+{\bf p})\,
Q(\omega, {\bf k},{\bf p})\;.
\end{align}
Here $\omega$ and ${\bf k}$ are the loop frequency and momentum, ${\bf
p}$ is the external momentum, $\Pi_{1,2}$ are the propagators of the
fields in the loop, and $Q$ is a polynomial in momenta and frequency
coming from the vertices. The propagators and vertices carry
tensor indices that we have suppressed. We have also
used that the external frequency vanishes as the background shift
$\bN_i$ was chosen to be time-independent. Similarly, the tadpole
diagrams (the last two diagrams in Fig.~\ref{fig:diagrams}) contain
integrals of the form,
\begin{align}
\label{tadInt}
\int \frac{\di\omega \di^3k}{(2\pi)^4}\,
\Pi(\omega,{\bf k})\, Q(\omega, {\bf k},{\bf p})\;.
\end{align}
The above integrals feature logarithmic divergences that we want to
extract. To this end, we Taylor expand the integrands in
(\ref{fishInt}), (\ref{tadInt}) and focus on the terms quadratic in
the external momentum $p_i$. Next we average over directions of the
loop momentum using the standard formulas specified to $d=3$ dimensions
\cite{Collins:1984xc},
\be
\overline{k_{i_1}k_{i_2}...k_{i_n}}= 
\begin{cases}
0~,& \text{if $n$ is odd}\\
\frac{k^n}{n+1} T_{i_1i_2...i_n}\;,& 
\text{if $n$ is even}
\end{cases}
\ee
where
\begin{align}
T_{i_1 i_2...i_n}=\frac{1}{n!}\left[ \delta_{i_1i_2}...\delta_{i_{n-1}i_n}+\text{all permutations of the i's}\right]\;.
\end{align}
In this way we are left with a sum of terms containing various tensor
structures multiplied by the integrals
\begin{align}
\label{genInt}
\int \frac{\di\omega \di^3 k}{(2\pi)^4}\, 
\omega^{2a} k^{2b} \big({\cal P}_I(\omega,k)\big)^A 
\big({\cal P}_J(\omega,k)\big)^B\;, 
\end{align}
with constant exponents $a,b,A,B$. Here ${\cal P}_{I,J}$, $I,J\in
\{tt,s,1,2\}$ are the pole structures listed in (\ref{poles}). Note
that at most two different pole structures can appear in an individual
integral. The exponents in (\ref{genInt}) satisfy the relation 
\be
\label{exponrel}
6+6a+2b-6A-6B=0\;,
\ee
so that the integrals are logarithmically divergent.

To regularize the UV divergences we use the Schwinger 
representation for each pole function, 
\begin{align}
\big({\cal P}_I(\omega,k)\big)^A=\int_0^\infty \frac{\di s_I
  s_I^{A-1}}{\Gamma(A)}
\,e^{-s_I ({\cal P}_{I}(\omega,q))^{-1}}\;,
\end{align}
where $\Gamma(z)$ is the gamma-function.
The integration over $\omega$ and ${\bf k}$ is then performed using
the formula,
\be
\int \frac{\di\omega \di^3 k}{(2\pi)^4}\, 
\omega^{2a} k^{2b}\;e^{-U\omega^2-Vk^6}=
\frac{1}{24\pi^3}\Gamma\left(a+\frac{1}{2}\right)
\Gamma\left(\frac{b}{3}+\frac{1}{2}\right)
U^{-a-\frac{1}{2}}\;V^{-\frac{b}{3}-\frac{1}{2}}\;.
\ee
Finally, if the resulting integral involves two Schwinger parameters,
we integrate over one of them. Namely, we write,
\be
\int_0^\infty \!\!\di s_1 \di s_2\;s_1^{A-1} s_2^{B-1}
(s_1+s_2)^{-a-\frac{1}{2}}
(s_1+u\,s_2)^{-\frac{b}{3}-\frac{1}{2}}
\!=\!\!\int_0^\infty\frac{\di s_1}{s_1}\!\int_0^\infty
\!\!\di z\, \frac{z^{B-1}}{(1+z)^{a+\frac{1}{2}}
(1+u\,z)^{\frac{b}{3}+\frac{1}{2}}}\,, 
\ee
where we made the change of variables $z=s_2/s_1$ and used the
relation (\ref{exponrel}). The integral over $z$ is convergent
and for all values of $a,b,B$ appearing in our calculation can be
taken in terms of elementary functions. The integral over the last
remaining Schwinger parameter is kept to the end of the calculation
and encodes the logarithmic divergence. \\

We have implemented the computational procedure described in this
section in a Mathematica code. Manipulation of tensor objects
was performed using the package {\it xAct} \footnote{\url{http://www.xact.es}}.
Some parts of the
computation were repeated using an alternative code written 
in
FORM\footnote{\url{https://www.nikhef.nl/~form/}}
\cite{Vermaseren:2000nd}. 
For a check of the procedure and the code we
also performed the calculations in the case of $(2+1)$ dimensions
where we reproduced the results of \cite{Barvinsky:2017kob}.

\section{The $\beta$-functions of $G$ and $\lambda$}
\label{sec:results}

Evaluation of the one-loop diagrams depicted in
Fig.~\ref{fig:diagrams} produces a divergent correction to the
effective Lagrangian for $\bN^i$,
\be
\label{L1loop}
\delta {\cal L}_{\bN}^{\rm 1-loop,\,div}
=\Big(C_1\, p_i \bN_j p_i \bN_j + C_2\, (p_i \bN_i)^2\Big)
\int \frac{\di s}{s}\;.
\ee
Comparing this with (\ref{eq:lagrangian_bg}) we extract the renormalized couplings,
\be
G_R=G-4G^2 C_1\int \frac{\di s}{s}~,~~~~~
\l_R=\l+2G\big((1-2\l)C_1-C_2)\int \frac{\di s}{s}\;.
\ee
We interpret the logarithmic divergence as the result of integrating
out high-momentum 
modes in the Wilsonian approach
to renormalization. Thus, we identify\footnote{The power of momentum
  under the logarithm 
  is determined by the scaling dimension of the Schwinger parameter,
  $[s]=-6$.}, 
\begin{align}\label{eq:regulator}
\int\frac{\di s}{s}\simeq \log\left(\frac{\Lambda^6_{\rm UV}}{k_*^6}\right)\;,
\end{align}
where $\Lambda_{\rm UV}$ is a UV cutoff for momentum 
and $k_*$ is the RG reference scale. $\b$-functions are defined as
sensitivities of the couplings to $k_*$, which yields,
\bseq
\begin{align}
\label{betaG}
&\beta_G\equiv\frac{dG_R}{d\log k_*}=24G^2 C_1\;,\\
\label{betalam}
&\beta_\l\equiv\frac{d\l_R}{d\log k_*}=12 G \big(C_2-(1-2\l)C_1\big)\;.
\end{align}
\eseq

The computation of the coefficients $C_{1,2}$ for general gauge
parameter $\{\sigma,\xi,\varsigma\}$ is rather cumbersome due to the
presence of four different pole structures in the propagators that
generate a large number of terms for every diagram. The situation is
simplified if one chooses the gauge parameters in such a way that some
of the pole structures coincide. In this section we fix
\be
\label{gaugea}
\xi=-\frac{1-2\l}{2(1-\l)}\;,
\ee 
which collapses the two poles corresponding to the gauge modes, ${\cal
P}_1={\cal P}_2$. In order to keep track of the
gauge (in)dependence of the result we allow the two remaining gauge
parameters $\sigma,\varsigma$ to take arbitrary values. Results
for two other choices of gauge are presented in Appendix~\ref{app:B}. 

In the gauge (\ref{gaugea}) we obtain,
\begin{align}
\b_G=&\frac{G^2 }{40 \pi ^2 \left(1+\sqrt{\alpha}\right)\sqrt{\a} 
(1-\lambda ) (1-3 \lambda) 
\sqrt{\nu_5}}\Big[
- \alpha (1-3 \lambda ) \big(5 (5-4 \lambda) 
\sqrt{2\sigma  \nu_5}+18-14 \lambda\big)\notag\\
&-\sqrt{\alpha} \left(5(1-3 
\lambda) (5-4 \lambda) \sqrt{2\sigma  \nu_5}
+53-142 \lambda+99 \lambda^2\right)
-27+(74-57 \lambda ) \lambda \Big]\;,
\label{betaGa}\\
\notag\\
\label{betalam1}
\beta_\l=&
\frac{27(1-\l)^2+3\sqrt{\a}(11-3\l)(1-\l)-2\a(1-3\l)^2}{
120\pi^2(1+\sqrt{\a})\sqrt{\a}(1-\l)}\frac{G}{\sqrt{\nu_5}}\;,
\end{align}
where we have introduced
\be
\label{alpha}
\alpha=\frac{\nu_s}{\nu_5}\;.
\ee
We observe that both expressions are independent of the gauge
parameter $\varsigma$. On the other hand, $\b_G$ depends on
$\sigma$. This is not unexpected. Indeed, a change of gauge is known
to affect the background effective action by adding to it a
contribution which is a linear combination of equations of motion
\cite{DeWitt:1967ub,Kallosh:1974yh}. 
Such terms vanish for on-shell background configurations, but not in
general. The background used in our calculations is not on-shell, so
the effective action evaluated on it is gauge-dependent.
In Appendix~\ref{app:C} we show that 
a change of gauge induces the following shift of the renormalized
couplings, 
\be
\label{gaugeshift}
G\mapsto G-2G^2\epsilon~,~~~~~
\l\mapsto\l~,~~~~~
\nu_a\mapsto\nu_a-4G\nu_a\epsilon\;,
\ee
where $\epsilon$ is an infinitesimal parameter.
We see that $G$, as well as $\nu_a$, transform non-trivially and
therefore their renormalization and $\b$-functions are not
gauge-invariant. On the other hand, $\l$ remains untouched by
(\ref{gaugeshift}). It represents an {\em essential} coupling of the
theory and its $\b$-function must be gauge-independent. This is indeed
born out by our computation, see (\ref{betalam1}). Other essential
couplings can be chosen, for example, as $G/\sqrt{\nu_5}$, $\a$ and
$\nu_a/\nu_5$, $a=1,2,3$.

\begin{figure}
\includegraphics[scale=.33]{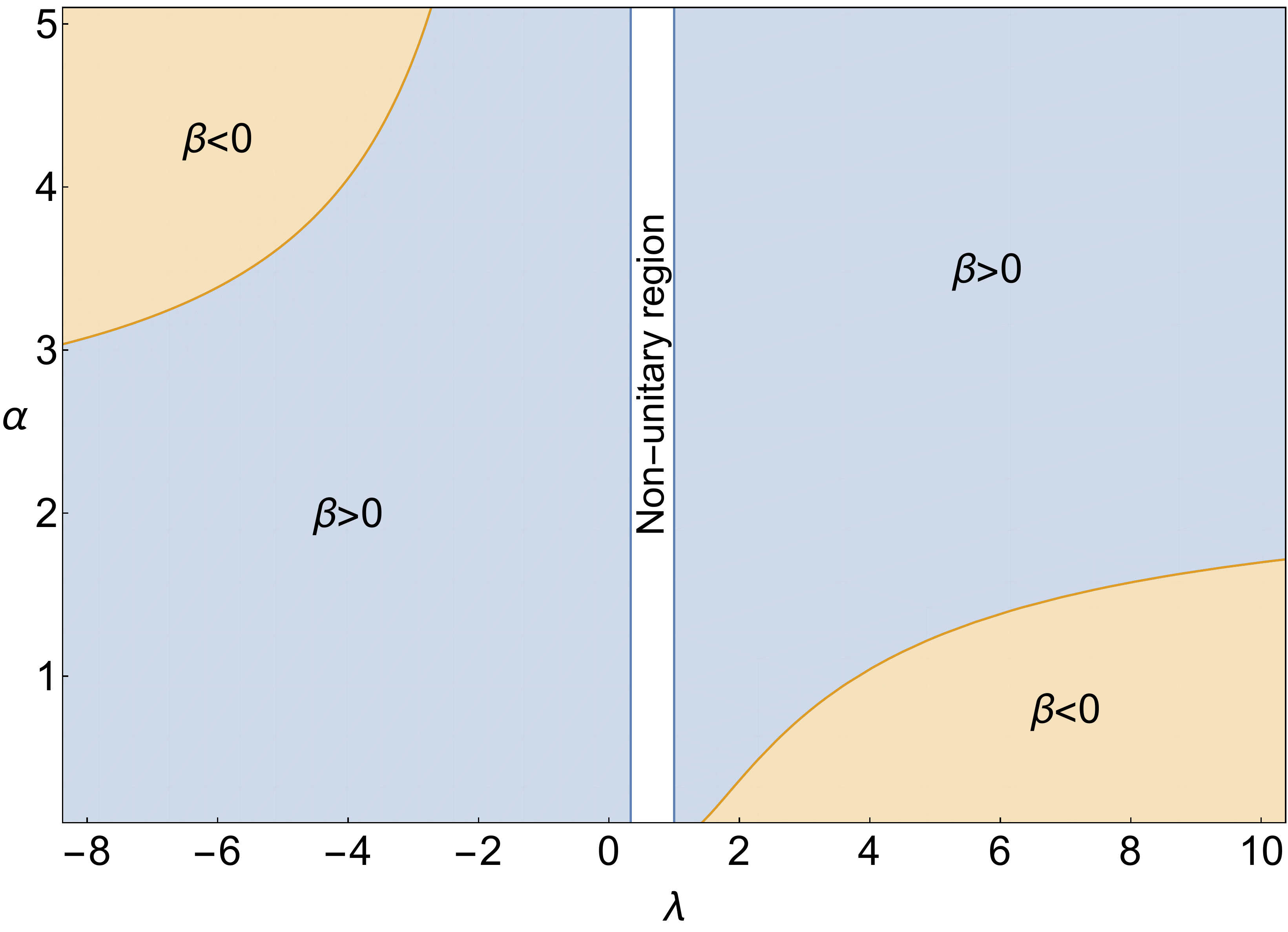} 
\caption{The sign of the $\beta$-function for the coupling $\l$ in 
different regions of parameter $(\lambda,\alpha)$. 
The $\b$-function vanishes on the lines separating positive and
negative regions. The line at $\l>1$ ($1/3<\l$) is UV attractive
(repulsive) along the $\l$-direction. These lines correspond to
potential location of fixed points of the full RG flow. The region
$\lambda \in [1/3,1]$ is excluded by the requirement of unitarity.}
\label{fig:plot}
\end{figure}

Let us discuss the properties of the gauge-invariant function $\b_\l$,
eq.~(\ref{betalam1}). Of particular interest is the manifold in the
parameter space of the theory where this function vanishes. It
constitutes a potential location of the fixed points of the full RG
flow. The contribution of $G/\sqrt{\nu_5}$ factors out, so that the
sign of $\b_\l$ and the location of its zeros are determined only by
two parameters $\l$ and $\a$. We have the following regions in the
plane of these parameters\footnote{Recall that the interval
  $1/3\leq\l\leq 1$ is excluded as it corresponds to negative kinetic
  term of the scalar mode and hence the loss of unitarity.},
see Fig.~\ref{fig:plot}:
\begin{itemize}
\item
$\l<1/3~,~~\a\leq 9/4$

In this region $\beta_\l$ is always positive. So it does not
contain any fixed points.

\item $\l<1/3~,~~\a> 9/4$

$\beta_\l$ is positive at $\l>\l_\a$ and negative at $\l<\l_\a$, where
\be
\label{lama}
\l_\a=\frac{9+7\sqrt{\a}-2\a+2\sqrt{10(\a+\a^{3/2})}}{3(3+\sqrt{\a}-2\a)}\;.
\ee
The $\b$-function vanishes on the line $(\l_\a,\a)$ which is IR
attractive along the $\l$-direction. 

\item
$\l>1~,~~\a< 9/4$

$\beta_\l$ is positive at $\l<\l_\a$ and negative at $\l>\l_\a$, where
$\l_\a$ is again given by (\ref{lama}). 
Now the line $(\l_\a,\a)$ is attractive along the $\l$-direction in
UV. When embedded into the full 6-dimensional space of the essential
couplings of the
theory, this line is promoted to a codimension-one surface. UV fixed
points of the full RG flow, if any, should lie on this surface.  

\item
$\l>1~,~~\a\geq 9/4$

Here $\beta_\l$ is always positive, and no fixed points
exist in this region.

\end{itemize}
Presence of UV attractive zeros of $\b_\l$ at finite values of 
$\l>1$ is analogous to the situation in $(2+1)$ dimensions, where such
zero is known to correspond to an asymptotically free UV fixed
point~\cite{Barvinsky:2017kob}. 

To carry the comparison with the $(2+1)$-dimensional case further, it
is instructive to work out the behavior of $\b_\l$ when $\l$
approaches the boundaries of the allowed regions, $\l\to 1/3^-$ or
$\l\to 1^+$, with all other couplings in the Lagrangian held fixed. As
implied by (\ref{dispscalar}), these limits correspond to degenerate
dispersion relation of the scalar graviton. One writes (see
eqs.~(\ref{nus}), (\ref{alpha})),
\be
\a=\frac{1-\l}{1-3\l}\bigg(\frac{8\nu_4}{\nu_5}+3\bigg)\;.
\ee 
Substituting this into the expression (\ref{betalam1}) one obtains,
\bseq
\begin{align}
&\b_\l\propto\sqrt{\frac{1}{3}-\l}& \text{at}~\l\to \frac{1}{3}^-\;,\\
&\b_\l\propto\frac{1}{\sqrt{\l-1}}& \text{at}~\l\to 1^+\;.
\end{align}
\eseq
In other words, the $\b$-function of $\l$ vanishes (diverges) 
as the square-root
when the frequency of the scalar graviton diverges (vanishes). 
This is the same qualitative behavior as in $(2+1)$
dimensions\footnote{In the $(2+1)$-dimensional case the boundaries of
  the allowed region correspond to $\l\to 1/2^-$, $\l\to 1^+$.}, 
see eq.~(22a) of~\cite{Barvinsky:2017kob}.

\section{Conclusions}
\label{sec:conclusions}
In this paper we have computed the renormalization group flow of $G$
and $\lambda$ in $(3+1)$-dimensional Ho\v rava Gravity. We used the
background-field method and exploited the gauge invariance of the
kinetic term which allowed us to focus on the renormalization of
operators constructed from the shift field. Setting the spatial
background metric to be flat we avoided major complications
arising due to the involved structure of the potential term.
We have considered a variety of gauge choices and found that the
$\beta$-function of $G$ depends on the gauge, as expected given that
this coupling is not {\it essential}, i.e. it cannot be defined using
only on-shell quantities. 

On the other hand, the coupling $\l$ is essential and its
$\b$-function given by eq.~(\ref{betalam1}) is gauge-independent. 
$\b_\l$ vanishes on two disjoint codimension-one surfaces 
in
the 6-dimensional parameter space of essential couplings. One of these
surfaces is UV attractive along the $\l$-direction, whereas the other
is UV repulsive. The former surface represents a potential location
for UV fixed points of the full RG flow. Of course, to make a
definitive conclusion about the existence of such fixed points one
has to calculate $\b$-functions of the remaining couplings of the
theory that enter in the potential part of the action. This
calculation will require other methods than the ones used in this paper and we
plan to report it elsewhere.

As a final remark, we note that the approach described in this
paper can be straightforwardly generalized to compute the running of
$G$ and $\l$ in any number of spacetime dimensions $(d+1)$. Indeed,
all interaction vertices used in our calculation come from the kinetic
term in the action and the part of the gauge-fixing term that contains
the shift field. These are universal. The dimension-dependent
potential entered in the calculation only via its quadratic expansion
that determined the propagators of the dynamical fields. This
quadratic part can be reconstructed for arbitrary $d$ from the
requirements of invariance under the Lifshitz scaling and the
linearized spatial diffeomorphisms and will be fully characterized by
the combination of couplings appearing in the dispersion relations of the
physical modes.

\section*{Acknowledgments}
We thank Diego Blas, Kevin Grosvenor, Ted Jacobson, Charles Melby-Thompson,
Christian Steinwachs and Ziqi Yan for useful discussions. 
This work was supported by the Swiss National Science Foundation and
the Russian Foundation for Basic Research grant No. 17-02-00651.

\appendix

\section{Interaction Lagrangian}
\label{app:Lint}

Here we present explicitly the parts of the interaction Lagrangian that are
relevant for the one-loop renormalization of the operators in the 
background action (\ref{eq:lagrangian_bg}).
These are obtained from the total
action (\ref{Stot}) and are of two types. First, there are cubic
interactions including one $\bar N^i$, with or without spatial
derivatives, and two dynamical fields:
\bseq
\label{Lintparts}
\begin{align}
\nonumber {\cal L}_{\bar{N}hh}=\frac{1}{2G}\bigg[&-\frac{1}{4}\bN_{ij} h\dot h_{ij}
-\frac{\l}{2}\bN_{ij}h_{ij}\dot h+\bN_k\d_i h_{jk}\dot h_{ij}-\frac{1}{2}\bN_k\d_k h_{ij}\dot h_{ij}
-\l\bN_k\d_ih_{ik}\dot h\\
&+\frac{\l}{2}\bN_k\d_k h \dot h
-\bN_k\d_lh_{li}\dot h_{ik}
+\l\bN_k \d_ih\dot h_{ik}\bigg],
\label{LintNhh}\\
{\cal L}_{\bar{N}hn}=\frac{1}{2G}\bigg[&-\bN_{ij}\d_i h_{jk}n_k
+\frac{1}{2}\bN_{ij} \d_kh_{ij} n_k
+\frac{1}{2}\bN_{ij}h\d_in_j-\bN_{ij}h_{ki}\d_k n_j
-\bN_{ij}h_{ik}\d_j n_k\notag\\
&+2\l \d_i\bN_i\d_j h_{jk} n_k -\l \d_i\bN_i\d_kh n_k
+2\l \d_i\bN_i h_{jk} \d_j n_k-\l \d_i\bN_i h \d_kn_k\notag\\
&+2\l \d_i\bN_j h_{ij}\d_kn_k
-\bN_k\d_i h_{jk} \d_i n_j-\bN_k \d_i h_{jk}\d_j n_i
+\bN_k \d_kh_{ij}\d_i n_j\notag\\
&+2\l\bN_k\d_lh_{lk}\d_in_i-\l\bN_k\d_k h \d_in_i
+\d_j\bN_i\d_l h_{li}n_j-\l\d_j\bN_i\d_ihn_j\notag\\
&-\bN_k\d_lh_{li}\d_kn_i+\l\bN_k\d_ih\d_kn_i
-\frac{\varsigma}{2}\bN_{ij}\d_k h_{ki}n_j
+\frac{\varsigma\l}{2}\bN_{ij}\d_ihn_j\bigg],\\
{\cal L}_{\bar{N}\pi h}=\frac{i}{G}\bN_k&\pi_i\dot h_{ik}\;,\\
{\cal L}_{\bN \pi n}=\frac{1}{2G}\big[&-2i\d_j\bN_i\pi_in_j
+2i \bN_k\pi_i \d_kn_i
+i\varsigma \bN_{ij}\pi_in_j\big]\;,\\
\label{LghNcc}
{\cal L}_{\bN\bar c c}=-\frac{1}{G}\bigg[&-2\bN_k\bar c_i\d_k\dot c_i
+2\d_k \bN_i\bar c_i\dot c_k-\frac{\varsigma}{2}\bN_{ij}\bar c_i\dot
c_j\bigg]\;,
\end{align}
\eseq
where $\bar N_{ij}\equiv \d_i \bN_j+\d_j\bN_i$ and in deriving
(\ref{LintNhh}) we eliminated some terms that combine into a total time
derivative\footnote{Expressions (\ref{Lintparts}) 
can be somewhat simplified by performing
  further integration by parts. We do not write the resulting
  expressions in order not to overload the paper with formulas.}. 
Second, we need the quartic interactions involving two $\bN_i$ and two
dynamical fields. In this case divergent contributions come from
tadpole 
diagrams where the two lines of the dynamical fields stemming from the
same vertex connect with each other to form a loop. Thus, to find the 
divergent contributions of the form
(\ref{eq:lagrangian_bg}), we can restrict only to those terms in the
quartic Lagrangian that contain exactly two spatial derivatives
acting on the $\bN_i$-legs. These have the form, 
\bseq
\label{LintNNparts}
\begin{align}
\nonumber \\
{\cal L}_{\bN\bN hh}=\frac{1}{2G}\bigg[&\bN_{ij}^2
\bigg(-\frac{1}{16}h_{kl}^2+\frac{h^2}{32}\bigg)
+\frac{1}{2}\bN_{ij}\bN_{kj}h_{il}h_{kl}
-\frac{1}{4}\bN_{ij}\bN_{kj}h_{ik}h
+\frac{1}{4}\bN_{ij}\bN_{kl}h_{ik}h_{jl}\notag\\
&-\!\l (\d_i\bN_i)^2\bigg(\!\!-\!\frac{1}{4}h_{kl}^2+\!\frac{h^2}{8}\bigg)
\!-\!\frac{\l}{4}\bN_{ij}\bN_{kl}h_{ij}h_{kl}
+\!\frac{\l}{2}\bN_{ij}\d_k \bN_k h_{ij} h
\!-\!\l \bN_{ij}\d_k\bN_k h_{il}h_{jl}\bigg],\\
\label{LghNNcc}
{\cal L}_{\bN \bN\bar c c}=-\frac{1}{G}\bigg[&
-\bN_k\d_k\d_l\bN_i\bar c_ic_l+\d_k\bN_i\d_l\bN_k\bar c_ic_l
-\frac{\varsigma}{2}\bN_{ij}\d_k\bN_j\bar c_i c_k\bigg]\;.
\end{align}
\eseq

\section{Alternative gauges}
\label{app:B}

In this appendix we report the results of the computation in two
gauges outside the family considered in Sec.~\ref{sec:results}. In
both cases the gauge parameters $\sigma,\xi$ are chosen in the way to
reduce the number of pole structures by making the gauge poles
coincide with the poles of the physical modes. The third parameter
$\varsigma$ is kept arbitrary.
\begin{itemize}
\item[i)] 
The first choice is
\be
\label{gaugec}
\s=\frac{1}{2\nu_5}~,~~~~~\xi=\frac{\nu_s}{2\nu_5(1-\l)}-1~~~~
\Longrightarrow~~~~
\P_1=\P_{tt},~\P_2=\P_s\;. 
\ee
It leads to the following $\b$-function for the coupling $G$,
\be
\label{betaGc}
\b_G=-\frac{32-89\l+57\l^2+3\sqrt{\a}(26-79\l+53\l^2)+2\a(19-74\l+51\l^2)}
{40\pi^2 (1+\sqrt{\a})\sqrt{\a}(1-\l)
  (1-3\l)}\frac{G^2}{\sqrt{\nu_5}}\;,
\ee
where $\a$ is defined in (\ref{alpha}).

\item[ii)]
The second choice,
\be
\label{gauged}
\s=\frac{1}{2\nu_s}~,~~~~~\xi=\frac{\nu_5}{2\nu_s(1-\l)}-1\;,
~~~~
\Longrightarrow~~~~
\P_1=\P_s,~\P_2=\P_{tt} \;,
\ee
yields
\be
\label{betaGd}
\b_G=-\frac{47-154\l+117\l^2+3\sqrt{\a}(26-79\l+53\l^2)+a(23-83\l+42\l^2)}
{40\pi^2 (1+\sqrt{\a})\sqrt{\a}(1-\l)
  (1-3\l)}
\frac{G^2}{\sqrt{\nu_5}}\;.
\ee
\end{itemize}
We observe that while $\b_G$ is independent of $\varsigma$, it is
different in the two gauges, as expected from the gauge-dependence of
the renormalized Lagrangian. 

On the other hand, we have found that the $\b$-function of $\l$ is the
same in both gauges (\ref{gaugec}), (\ref{gauged}) and coincides with
the expression (\ref{betalam1}) from the main text. 
Thus gauge-independence provides a further check of our calculational
procedure.

\section{Ambiguity of the one-loop effective action}
\label{app:C}

It is well-known that in gauge theories the background effective
action $\Gamma_{\rm eff}$ depends on the choice of gauge fixing for
the fluctuating fields, unless the background satisfies the equations
of motion. This dependence is expressed by the statement that a change
in the gauge fixing leads to the shift of the effective action,
\be
\label{Geffshift}
\Gamma_{\rm eff}\mapsto \Gamma_{\rm eff}+\epsilon {\cal A}\;,
\ee
where $\epsilon$ is an infinitesimal parameter and ${\cal A}$ is a
functional of the background fields which is a linear combination of
the equations of motion \cite{DeWitt:1967ub,Kallosh:1974yh}.
Our task is to see if such a functional exists in the projectable HG.
To be in line with the main
text, we work in Euclidean time and focus on the high-energy
part of the action given by (\ref{eq:action}).

As we are interested only in the one-loop divergence of $\Gamma_{\rm
  eff}$ which is an integral of a local Lagragian density, the
functional ${\cal A}$ must also have this form. Moreover, it should be
invariant under FDiff and contain only marginal operators with respect
to the Lifshitz scaling. Variation of (\ref{eq:action}) with respect
to $N^i$ and $\g_{ij}$ produces the equations,
\bseq
\begin{align}
\label{eqN}
&\nabla_j K^j_i-\l \nabla_i K=0\;,
\\
&-\!\frac{\d}{\d \t}\big[\sqrt{\g}\big(K^{ij}\!-\!\l\g^{ij} K\big)\big]
\!-\!2\sqrt{\g}\big(K^i_k K^{kj}\!-\!\l K^{ij} K\big)
\!+\!\frac{1}{2} \sqrt{\g}\,\g^{ij} \big(K_{kl} K^{kl}\!-\!\l K^2\big)
\!+\!\frac{\delta (\sqrt{\g}\,{\cal V})}{\delta\g_{ij}}=0\,,
\label{eqgamma}
\end{align}
\eseq
where for simplicity we set $N^i$ to zero by a choice of gauge. Wo do
not write explicitly the result of the variation of the
potential term which is too lengthy. The l.h.s. of the first equation
is a vector operator of dimension 4. Any other local covariant vector
operator in the theory has at least dimension 3 (this is the case for
$\nabla_iR$). Thus, we cannot use the l.h.s. of (\ref{eqN}) to
construct a local Lagrangian density of dimension 6.

On the other hand, the l.h.s. of (\ref{eqgamma}) already has dimension
6. To get a scalar density under FDiff we contract it with 
$\g_{ij}$. This yields,
\be
\label{traceeq}
-(1-3\l)\frac{\d(\sqrt{\g}K)}{\d \t}
+\frac{3}{2}\sqrt{\g}\big(K_{ij}K^{ij}-\l K^2\big)
+\g_{ij}\frac{\delta (\sqrt{\g}\,{\cal V})}{\delta\g_{ij}}=0\;.
\ee
The last term on the l.h.s. represents the result of the variation of the
potential part with respect to spatial Weyl transformations,
$\g_{ij}\mapsto \Omega\g_{ij}$. By inspection of the expression
(\ref{eq:action}), we find that for Weyl rescalings with a
space-independent factor $\Omega$ the potential part transforms as
$\sqrt{\g}\,{\cal V}\mapsto \Omega^{-3/2}\sqrt{\g}\,{\cal V}$. This
implies that for a general rescaling the variation of
$\sqrt{\g}\,{\cal V}$  has the form,
\be
\g_{ij}\frac{\delta (\sqrt{\g}\,{\cal V})}{\delta\g_{ij}}
=-\frac{3}{2} \sqrt{g}\,{\cal V}+(\text{total spatial derivatives})\;.
\ee
Therefore, the l.h.s. of (\ref{traceeq}) can be written as
\be
\frac{3}{2}\sqrt{\g}\big(K_{ij}K^{ij}-\l K^2-{\cal V}\big)
+(\text{total derivatives})\;.
\ee
Integrating over the whole spacetime and neglecting total derivatives
one obtains,
\be
\label{addaction}
{\cal A}=\int \di \t\di^3x\sqrt{\g}\,\Big(K_{ij}K^{ij}-\l K^2-{\cal V}\Big)\;.
\ee
Adding it to the effective action as in (\ref{Geffshift}) induces a
shift of the couplings (\ref{gaugeshift}) discussed in the main text. 

Let us mention another way to arrive to the expression
(\ref{addaction}). Namely, one can restore
time reparameterizations by reintroducing in  (\ref{eq:action}) the
lapse function $N(t)$. Then variation with respect to $N$ produces a
global Hamiltonian constraint\footnote{Note the negative sign in front
of the potential term which is the consequence of working in
Euclidean 
time.},
\be
\int \di^3x\sqrt{\g}\,\big(K_{ij}K^{ij}-\l K^2-{\cal V}\big)=0\;.
\ee
Clearly, setting $N$ back to 1 and integrating the l.h.s. of this
equation over time we recover~(\ref{addaction}).

\bibliography{lambda31}
\bibliographystyle{apsrev4-1}
\end{document}